\documentclass[mathleft]{an}
\usepackage{graphicx}
\usepackage{times}
\begin{document}
\sloppy

\newcommand{\pd}{\partial}

\title{Solar dynamo models with $\alpha$-effect and turbulent pumping
  from local 3D convection calculations}

\author{P. J. K\"apyl\"a\inst{1,2,3}\fnmsep\thanks{Corresponding author:
  \email{petri.kapyla@oulu.fi}\newline}, M.J. Korpi \inst{1} \and I. Tuominen \inst{1}}

\titlerunning{Solar dynamo models with transport coefficients from local
 convection calculations}
\authorrunning{P.J. K\"apyl\"a et al.}

\institute{ 
Observatory, University of Helsinki, PO BOX 14, FI-00014 University of 
Helsinki, Finland
\and
Astronomy Division, Department of Physical Sciences, PO BOX 3000,
FI-90014 University of Oulu, Finland
\and
Kiepenheuer-Institut f\"ur Sonnenphysik, Sch\"oneckstrasse 6, D-79104
Freiburg, Germany}

\date{Received $<$date$>$; 
accepted $<$date$>$;
published online $<$date$>$}


\abstract{Results from kinematic solar dynamo models employing
  $\alpha$-effect and turbulent pumping from local convection
  calculations are presented. We estimate the magnitude of these
  effects to be around 2--3\,m\,s$^{-1}$, having scaled the local
  quantities with the convective velocity at the bottom of the
  convection zone from a solar mixing-length model. Rotation profile
  of the Sun as obtained from helioseismology is applied in the
  models; we also investigate the effects of the observed surface
  shear layer on the dynamo solutions. With these choices of the
  small- and large-scale velocity fields, we obtain estimate of the
  ratio of the the two induction effects, $C_\alpha/C_\Omega \approx
  10^{-3}$, which we keep fixed in all models. We also include a
  one-cell meridional circulation pattern having a magnitude of
  10--20\,m\,s$^{-1}$ near the surface and 1--2\,m\,s$^{-1}$ at the
  bottom of the convection zone. The model essentially represents a
  distributed turbulent dynamo, as the $\alpha$-effect is nonzero
  throughout the convection zone, although it concentrates near the
  bottom of the convection zone obtaining a maximum around $30\degr$
  of latitude. Turbulent pumping of the mean fields is predominantly
  down- and equatorward. The anisotropies in the turbulent diffusivity
  are neglected apart from the fact that the diffusivity is
  significantly reduced in the overshoot region. We find that, when
  all these effects are included in the model, it is possible to
  correctly reproduce many features of the solar activity cycle,
  namely the correct equatorward migration at low latitudes and the
  polar branch at high latitudes, and the observed negative sign of
  $B_r B_{\phi}$. Although the activity clearly shifts towards the
  equator in comparison to previous models due to the combined action
  of the $\alpha$-effect peaking at midlatitudes, meridional
  circulation and latitudinal pumping, most of the activity still
  occurs at too high latitudes (between $5\degr \ldots
  60\degr$). Other problems include the relatively narrow parameter
  space within which the preferred solution is dipolar (A0), and the
  somewhat too short cycle lengths of the solar-type solutions. The
  role of the surface shear layer is found to be important only in the
  case where the $\alpha$-effect has an appreciable magnitude near the
  surface.}

\maketitle

\section{Introduction}
The early solar dynamo models, initiated by Parker
(\cite{Parker1955}), relied on a scalar $\alpha$-effect and a
negative gradient of the angular velocity, $\Omega$, in the convection
zone (e.g. Steenbeck \& Krause \cite{SteenKrau1969}; Deinzer \& Stix
\cite{DeinzerStix1971}; K\"ohler \cite{Koehler1973}). These simple
models were enormously succesful in reproducing many of the observed
features of the solar cycle, most prominently the equatorward
propagating activity belts.

During the recent years, however, helioseismology has revealed that
the radial differential rotation shows a positive gradient near the
equator, whilst a negative gradient occurs only at high
latitudes. Moreover, the strongest gradients occur near the poles and
at the bottom of the convection zone (e.g. Schou et
al. \cite{Schouea1998}; Thompson et al. \cite{Thompsonea2003}). The
sign of the $\Omega$-gradient, if combined with positive (negative)
$\alpha$-effect in the Northern (Southern) hemisphere, leads to
poleward migration of the activity belts near the equator and to
equatorward migration in the polar regions according to mean-field
dynamo theory (Parker \cite{Parker1955}; Yoshimura
\cite{Yoshimura1975}). This is in contradiction with the observed
migration of solar activity tracers, being equatorward at low
latitudes, whilst a weaker polarward wave occurs at high
latitudes. The observed strong differential rotation in the polar
regions, together with the commonly adopted description of the
$\alpha$-effect being proportional to $\cos{\theta}$ therefore peaking
at the poles, leads to dynamo solutions with high-latitude activity;
virtually no sunspots, however, are observed above the latitude $\pm
40\degr$.

A suitable meridional circulation pattern is one possible remedy to
the aforementioned problems. At the moment it is known that near the
surface of the Sun, i.e. the topmost 10--20\,Mm, there is a poleward
flow of the order of 10--20\,m\,s$^{-1}$ (Zhao \& Kosovichev
\cite{ZhaoKosov2004}; Komm et al. \cite{Kommea2004}), but very little
is known about the structure of the flow in deeper layers. Models of
the solar differential rotation predict that there should be a return
flow near the bottom of the convection zone with a magnitude of
approximately $1$\,m\,s$^{-1}$ (e.g. Kitchatinov \& R\"udiger
\cite{KitchRued1999}; Rempel \cite{Rempel2005}). The meridional flow
structure and a low value of the turbulent diffusivity, $\eta_{\rm
  t}$, are of critical importance for the so-called flux transport
dynamos (e.g. Choudhuri, Sch\"ussler \& Dikpati
\cite{Choudhuriea1995}; Dikpati \& Charbonneau
\cite{DikChar1999}). These models are essentially $\alpha
\Omega$-dynamos where the meridional flow works as a conveyor belt for
the poloidal magnetic field, and the $\alpha$-effect is only due to
the decay of active regions near the solar surface, or due to
instabilities in the magnetic layer below the convection zone
(e.g. Schmitt, Sch\"ussler \& Ferriz-Mas \cite{Schmittea1996};
Brandenburg \& Schmitt \cite{BranSchm1998}; Dikpati \& Gilman
\cite{DikGil2001}), whereas the shear in the tachocline is responsible
for the generation of the toroidal field. However, in the present
paper we do not exclude the possibility to have an $\alpha$-effect due
to cyclonic turbulent motions in the convection zone proper as is
expected in the mean-field theory and suggested by the local
convection calculations (e.g. Brandenburg et al. \cite{Brandea1990};
Ossendrijver, Stix \& Brandenburg \cite{Osseea2001}, Ossendrijver et
al. \cite{Osseea2002}; K\"apyl\"a et al. \cite{Kaepylaeea2006},
hereafter KKOS). In this case it is questionable whether the
meridional flow alone is able to counter the missing activity and
wrong migration of the activity belts near the equator.

Another possibility to solve the problems outlined above arises in the
form of dynamo coefficients. Most of the solar models have up to date
relied on a scalar $\alpha$-effect, or they have taken into account
a limited amount of the anisotropies, such as vertical pumping of the
mean magnetic field (e.g. Brandenburg, Moss \& Tuominen
\cite{BranMoTuo1992}), or the contributions to the $\alpha$-effect due
to the steeply decreasing turbulence intensity towards the bottom of
the convection zone (R\"udiger \& Brandenburg
\cite{RuedBran1995}). Recently, Ossendrijver et
al. (\cite{Osseea2002}) computed all of the coefficients $a_{ij}$ that
appear in the expansion
\begin{eqnarray}
\mathcal{E}_i = a_{ij} B_{j} + b_{ijk}\frac{\pd B_j}{\pd
  x_k} + \cdots \;, \label{equ:emf}
\end{eqnarray}
describing the small-scale effects on the large scales from local
numerical 3D models of stellar convection. Taking only the first
  term on the rhs into account, $\vec{\mathcal{E}}$ can be written as
\begin{eqnarray}
\vec{\mathcal{E}} = \overline{\vec{\alpha}} \vec{B} + \vec{\gamma}
\times \vec{B}\;,
\end{eqnarray}
where $\overline{\vec{\alpha}}$ and $\vec{\gamma}$ are defined via
\begin{eqnarray} 
\alpha_{ij} & = & \frac{1}{2}(a_{ij} + a_{ji})\;, \label{equ:alpha}\\
\gamma_i & = & -\frac{1}{2} \varepsilon_{ijk}a_{jk}\;, \label{equ:gamma}
\end{eqnarray}
and consist of the symmetric and anti-symmetric parts of $a_{ij}$. The
diagonal components of $\alpha_{ij}$ describe the generation of a large
scale magnetic field through the $\alpha$-effect. The vector
$\vec{\gamma}$ describes turbulent pumping of the mean field with a
velocity that is common for all field components. The off-diagonal
components of $\alpha_{ij}$ contribute to the field direction
dependent part of the pumping effect (see e.g. Ossendrijver et
al. \cite{Osseea2002}; KKOS).

The main findings for the $\alpha$-effect in the study of Ossendrijver
et al. (\cite{Osseea2002}) were that whereas it is highly anisotropic,
the component corresponding to $\alpha_{\phi \phi}$ is almost constant
in magnitude as a function of latitude for Coriolis number of ${\rm
  Co} = 2\,\Omega \tau \approx 2.5$, where $\tau$ is an estimate of
the convective turnover time. It was also established numerically that
the turbulent pumping of the mean magnetic field depends upon the
field component. The results also clearly indicate down- and
equatorward pumping of the mean magnetic fields that can alleviate the
problems facing solar dynamo models. Furthermore, recently KKOS
extended the study of Ossendrijver et al. (\cite{Osseea2002}) to the
rapid rotation regime, i.e. ${\rm Co} \approx 10$, corresponding to
the deep layers of the solar convection zone. The main result of this
study is that $\alpha_{\phi \phi}$, responsible for the generation of
the poloical field from the toroidal one, no longer peaks at the poles
but rather at around latitude $\Theta = 30\degr$, which could be of
further help for the solar dynamo models employing the helioseismic
rotation profile. The vertical pumping of the magnetic field is
downward near the poles, but can be upward near the equator. The
latitudinal pumping is always towards the equator, but it is
concentrated in a rather narrow latitude range, i.e. $|\Theta| <
30\degr$ near the equator. The vertical pumping of the toroidal field
was found to be very small or upward at latitudes $|\Theta| <
45\degr$.

In the present study we explore the implications of the local 3D
results for the solar dynamo by means of axisymmetric mean-field
dynamo models employing the observed internal rotation of the
Sun. Starting from a simple $\alpha$-profile ($\cos \theta$ in
latitude and constant in radius within the convection zone) we
demonstrate the problems facing solar dynamo models. Then we introduce
an $\alpha$-profile that captures the essentials of the magnitude and
latitude dependence found in local convection calculations. Then we
proceed by adding the pumping effects and finally a reasonable
meridional flow constrained by the flow velocity observed on the solar
surface. We also discuss the stability of quadrupolar and dipolar
field configurations for the chosen $\Omega$- and $\alpha$-profiles
with and without turbulent pumping and meridional flow. Finally, we
discuss briefly the phase dilemma of the radial and toroidal magnetic
fields which has been widely used as a constraint for solar dynamo
models. The remainder of the paper is organised as follows: in
Sect.~\ref{sec:model} the mean-field model is described in detail, and
in Sects.~\ref{sec:results} and \ref{sec:conclusions} we present the
results and the conclusions of the study.

\section{The model} 
\label{sec:model}
We study axisymmetric kinematic mean-field models of the solar dynamo
in spherical polar coordinates within a shell $0.6 R_\odot \leq r \leq
R_\odot$. The inner region of the shell up to $r_{\rm c} = 0.7
R_\odot$ models an overshoot region below the convection zone. We
solve the mean-field induction equation
\begin{eqnarray} 
\frac{\pd \vec{B}}{\pd t} &=& \nabla \times [(\vec{U} + \vec{\gamma})
  \times \vec{B} + \overline{\vec{\alpha}} \vec{B} - \eta_t \nabla
  \times \vec{B}]\;, \label{equ:vB}
\end{eqnarray} 
where $\vec{U}=\vec{\Omega} \times \vec{r}+\vec{U}^{\rm mer}$ is the
velocity, $\vec{r} = r \hat{\vec{e}}_r$ the radius vector,
$\vec{\Omega}=\Omega(r,\theta) \hat{\vec{k}}$ the prescribed rotation
profile, $\hat{\vec{k}}$ the unit vector along the rotation axis, and
$\vec{U}^{\rm mer} = (U_r,U_\theta,0)$ the prescribed meridional
flow. $\overline{\vec{\alpha}}$ and $\vec{\gamma}$ are given by
Eqs.~(\ref{equ:alpha}) and (\ref{equ:gamma}), respectively.
$\eta_{\rm t}$ is the turbulent diffusion which we treat as a scalar
field neglecting its tensorial nature for the time being. However, we
take into account the decrease of $\eta_{\rm t}$ in the overshoot
region via
\begin{eqnarray} 
\eta_{\rm t}(r) = \eta_{\rm c} + 0.45 \eta_0 \bigg[ 1 + \tanh \bigg(
  \frac{r - r_{\rm c}}{d_1} \bigg) \bigg]\;,
\end{eqnarray} 
where $\eta_{\rm c} = 0.1 \eta_0$, $d_1 = 0.015$, and $\eta_0$ the
value in the convection zone.

Under the assumption of axisymmetry, the magnetic field can be
represented with two scalar fields $A$ and $B$ according to
\begin{equation}
\vec{B}=\vec{B}_{\rm T} + \vec{B}_{\rm P} = \nabla \times (A
\hat{\vec{e}}_\phi) + B \hat{\vec{e}}_\phi\;.
\end{equation} 
Thus we obtain separate evolution equations for $A$ and $B$, which we
non-dimensionalize by choosing the following units:
\begin{eqnarray}
&[r]& = R_{\odot} \equiv R, \ \ [t]=R^2/\eta_0 \equiv \tau, \ \ [U]=\eta_0/R, \nonumber \\ 
&[\eta_t]& = \eta_0, \ \ [B] = B_0, \ \ [A]=R B_0, [\Omega]=\Omega_0,
\end{eqnarray}
where the subscript $0$ refers to a typical value of the quantity in
question. The resulting equations are given by Eqs.~(\ref{equ:A}) and
(\ref{equ:B}) in Appendix A. The models are controlled by the
dimensionless parameters
\begin{equation}
C_\alpha=\frac{\alpha_0 R}{\eta_0}, \ \ C_{\Omega}=\frac{\Omega_0
  R^2}{\eta_0}, \ \ C_U=\frac{u_0 R}{\eta_0}\;, \label{equ:Ci}
\end{equation}
describing the magnitude of the $\alpha$-effect, differential rotation
and meridional flow with respect to diffusion. In what follows,
$\alpha_0$ and $u_0$ are the maximum values of the $\alpha$-effect and
the latitudinal component of the meridional flow. 

The nonlinear back reaction of the magnetic field on the turbulence
and thus on the $\alpha$-effect is modelled via a simple
$\alpha$-quenching formula
\begin{equation}
\alpha_{ij}[r,\theta,\vec{B}(r,\theta)] =
\frac{\alpha_{ij}}{1 + [\vec{B}(r,\theta)/B_{\rm eq}]^2},
\end{equation}
where $B_{\rm eq}$ is the equipartition strength of the large-scale
magnetic field. In practise the value of $B_{\rm eq}$ determines the
saturation strength of the magnetic field and we set simply $B_{\rm
  eq}/B_0 = 1$. The actual equipartition value can be estimated from
typical values of the local turbulent velocity and density using a
mixing length model of the solar convection zone. The values $u_{\rm
  rms} \approx 10$\,m\,s$^{-1}$ and $\rho = 100$\,kg\,m$^{-3}$
corresponding to the situation at the bottom of the convection zone
yield saturation strengths of the order of a few kG, and similar
values are obtained throughout the convection zone. The same quenching
formula is also applied on the turbulent pumping effects.

The equations (\ref{equ:A}) and (\ref{equ:B}) are solved over the
radius $0.6R \leq r \leq R$ and colatitude $0 \leq \theta \leq \pi$
using a two-dimensional equidistant grid of $N_r \times N_\theta$
gridpoints. The boundary condition at the poles reads $A = B = 0$, and
in the radial direction pseudo-vacuum conditions
\begin{equation}
B = \frac{\pd (r A)}{\pd r}=0\;,
\end{equation}
are used at the surface, and perfectly conducting condition at the
bottom of the convection zone, i.e.
\begin{equation}
A = \eta_{\rm t} \frac{\pd (r B)}{\pd r} - \alpha_{\theta \theta} r
\frac{\pd A}{\pd r}=0\;.
\end{equation}
The numerical resolution is varied between $41 \times 81$ and $61
\times 121$ grid points. The constant timesteps used are $\Delta t =
2.5 \cdot 10^{-5}$ for the low resolution case and $\Delta t = 1.5
\cdot 10^{-5}$ for the higher resolution case. The code has been
validated using the ``dynamo benchmark" test cases (Arlt et
al. \cite{Arltea2006}); the pseudo-vacuum boundary conditions used
here were found to result in slightly reduced critical dynamo numbers
in the case of $\alpha \Omega$ dynamos if compared to methods with
real vacuum boundaries, but otherwise good agreement with other
methods was found. Due to this we prefer to keep the pseudo-vacuum
boundaries at the surface due to the simplicity of their
implementation. The code employs second order accurate explicit
spatial discretisation and first (Euler) or second order
(Adams-Bashforth-Moulton predictor-corrector; see e.g. Caunt \& Korpi
\cite{CK}) accurate time stepping.

\begin{figure}
\resizebox{\hsize}{!}
{\includegraphics{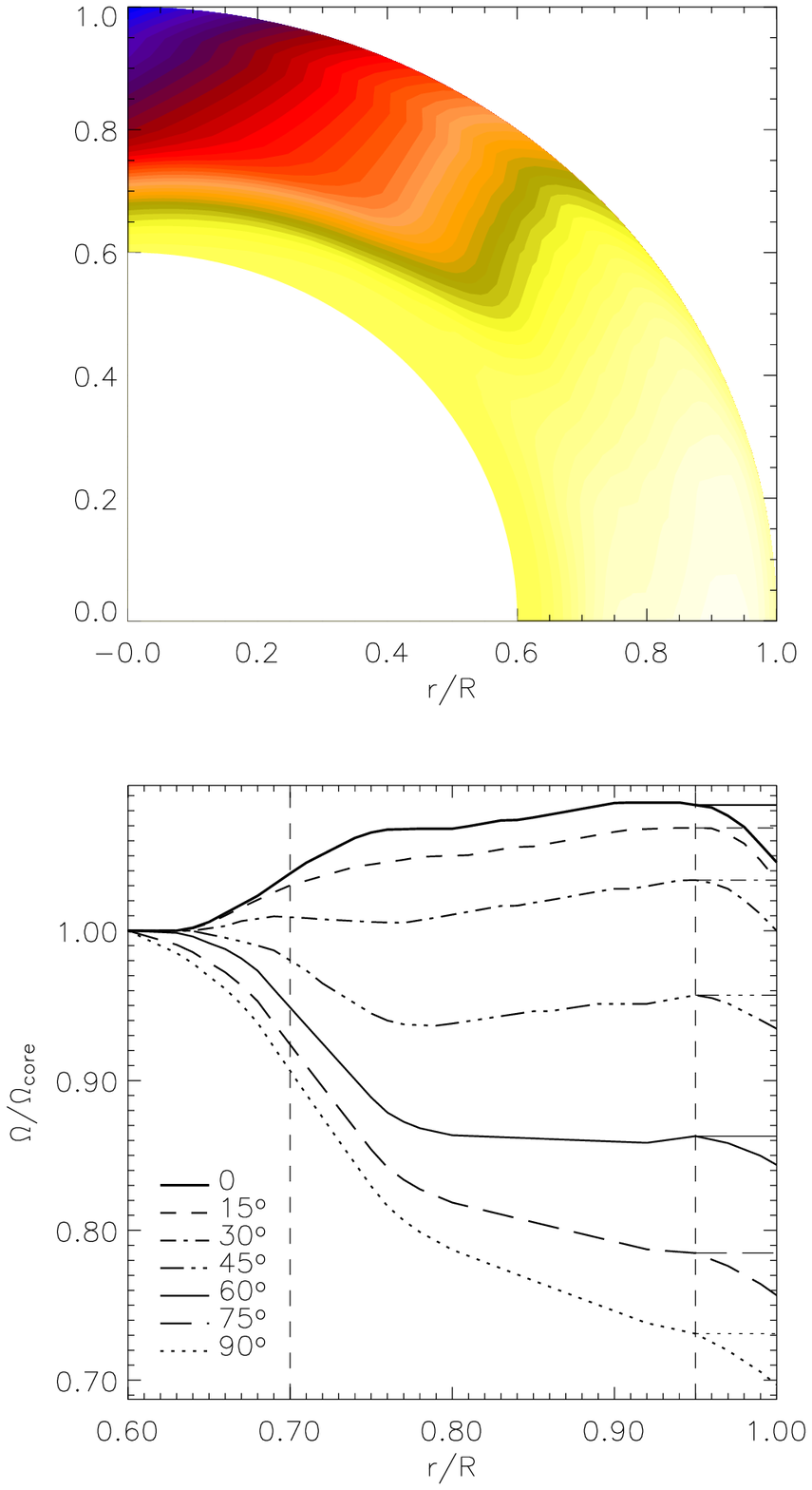}}
\caption{Upper panel: rotation profile adapted from the helioseismic
  results. The lower panel shows the angular velocity $\Omega$
  normalised by the core angular velocity, $\Omega_{\rm core} =
  430$\,nHz, at latitudes $0 \degr$ (equator, thick solid line), $15
  \degr$ (dashed), $30 \degr$ (dot-dashed), $45 \degr$
  (triple-dot-dashed), $60 \degr$ (solid), $75 \degr$ (long-dashed),
  and $90 \degr$ (pole, dotted line). The dashed vertical lines at $r
  = 0.70R$ and $r = 0.95R$ denote the bottom of the convection zone
  and the depth above which the surface shear layer is located,
  respectively. The thin horizontal lines above $r = 0.95R$ give
  $\Omega$ in the case where the surface shear is neglected.}
\label{fig:rotproheli}
\end{figure}

The $\overline{\vec{\alpha}}$-tensor and $\vec{\gamma}$-vector
component profiles are adapted from the local convection model of KKOS
(see below). The Coriolis numbers realised in the convection models
are here interpreted as depths in the convection zone (see Fig.~1 of
K\"apyl\"a et al. \cite{Kaepylaeea2005}), so that Co = 1 corresponds
to a depth of roughly 50 Mm below the surface, Co = 4 matches the
middle, and Co = 10 roughly the bottom of the solar convection
zone. We neglect the possibility to have an $\alpha$-effect in the
overshoot region in the present study. The latitudinal dependences are
approximative fits to the local calculations, made at four different
latitudes for the sets Co = 1 and Co = 4 with $30\degr$ steps, and
seven different latitudes for the set Co = 10 with $15\degr$
steps. The rotation profile used throughout the investigation is a
digitisation of the helioseismological inversion (e.g. Schou et
al. \cite{Schouea1998}), see Fig. \ref{fig:rotproheli}. Since no
reliable inversions are available for the polar regions, we assume the
profiles of $\Omega$ at $|\Theta| = 75\degr$ and $|\Theta| = 90\degr$
to be proportional to the profile at $60\degr$ via
\begin{eqnarray}
\Omega(r,75\degr) & = & \Omega(r,60\degr) \bigg[ 0.05 \Big( \frac{r_0 - r}{r_0 - R} \Big) - 1 \bigg]^2\;, \\
\Omega(r,90\degr) & = & \Omega(r,60\degr) \bigg[ 0.1 \Big( \frac{r_0 - r}{r_0 - R} \Big) - 1 \bigg]^2\;,
\end{eqnarray}
where $r_0 = 0.6$, and $\Theta = 90\degr - \theta$ is the
latitude. With this choice $\Omega$ at the surface on the poles is
consistent with the observed surface value.

In order to characterise the resulting solutions of the models in
terms of the large scale magnetic field structure, we define the
parity as
\begin{equation}
P = \frac{E^{(S)} - E^{(A)}}{E^{(S)} + E^{(A)}}\;, \label{equ:pari}
\end{equation}
where $E^{(S)}$ and $E^{(A)}$ are the energies of the symmetric and
antisymmetric parts of the field, respectively (see, e.g. Brandenburg
et al. \cite{Brandea1989}). For a symmetric quadrupolar solution (S0)
the parity is $+1$, and for an antisymmetric dipolar solution (A0),
$P=-1$.

\section{Results}
\label{sec:results}

\subsection{Estimates of $C_{\alpha}$ and $C_{\Omega}$ for the Sun}
Using the profiles of the transport coefficients from theoretical
considerations or local 3D convection calculations, the remaining task
is to choose appropriate values for the three dimensionless control
parameters $C_\alpha$, $C_\Omega$, and $C_U$, defined via
Eq.~(\ref{equ:Ci}). The local convection calculations of KKOS yield
${\rm max}(|\alpha_{ij}|)/u_{\rm rms} \approx 0.2$. We scale this to
physical units by using the mixing length estimate of the turbulent
velocity at the bottom of the convection zone, i.e. $u_{\rm rms}^{\rm
  MLT} \approx 10$\,m\,s$^{-1}$, and obtain a magnitude of
2\,m\,s$^{-1}$ for $\alpha_0$. Furthermore, using the mean angular
velocity of the Sun, $\Omega_0 = 2.6 \times 10^{-6}$\,s$^{-1}$, the
ratio $C_\alpha/C_\Omega = \alpha_0/(\Omega_0 R)$ is therefore
approximately $10^{-3}$; in the dynamo models describing the Sun, we
keep this ratio fixed. Furthermore, the meridional flow at the
surface, $u_0 \approx 10 \ldots 20$\,m\,s$^{-1}$, is known from
observations, from which we obtain another useful ratio for the solar
models, i.e. $C_U/C_\Omega = u_0/(\Omega_0 R) \approx 0.005 \ldots
0.01$.

The value of $\eta_0$ still remains a free parameter. Simple mixing
length estimates for the deep layers of the solar convection zone
(e.g. Stix \cite{Stix2002}) give $\eta_{\rm t} = \frac{1}{3} u_{\rm
  rms}^{\rm MLT} d \approx 5 \cdot 10^8$\,m$^2$\,s$^{-1}$, where $d =
\alpha_{\rm MLT} H_{\rm p}$, $\alpha_{\rm MLT} = 1.66$, and $H_{\rm p}
\approx 5 \cdot 10^7$\,m. Similar or slightly higher values are
obtained for all depths in the solar convection zone (see
e.g. Brandenburg \& Tuominen \cite{BrandTuo1988}). We find that
growing solutions are possible when $\eta_0 \approx
10^8$\,m$^2$\,s$^{-1}$ (see below).

\subsection{Constant $\alpha$ with $\cos{\theta}$ latitude dependence} 
\label{subsec:simpalpha}
\begin{figure}
\resizebox{\hsize}{!}
{\includegraphics{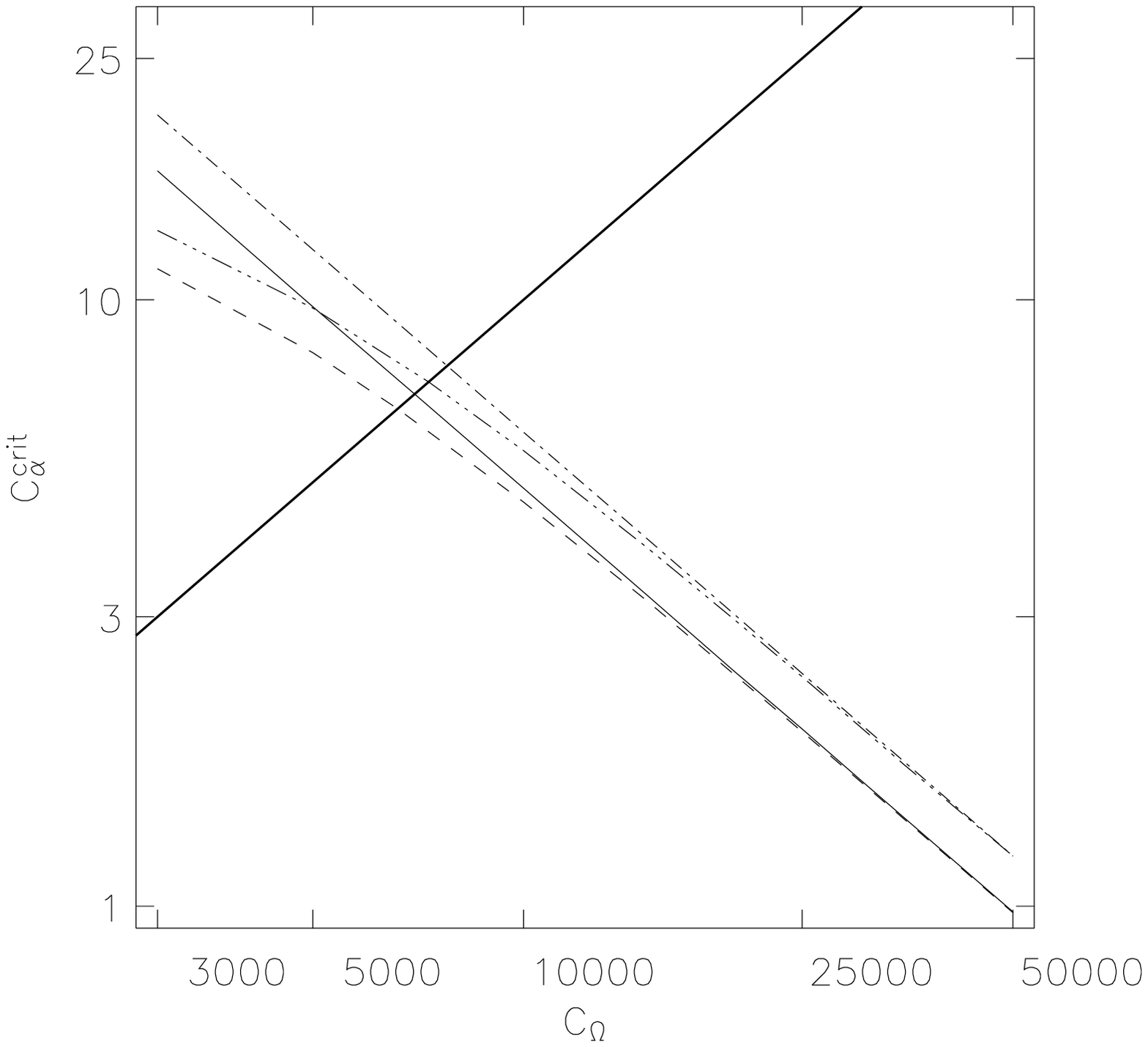}}
\caption{Critical dynamo numbers $C_\alpha^{\rm crit}$ as function of
  $C_\Omega$ for the $\alpha$-profile given by
  Eq.~(\ref{equ:simpalpha}). The solid and dashed lines give the
  results for the rotation profile including the surface shear, shown
  in Fig. \ref{fig:rotproheli}, and using the $\alpha \Omega$ and
  $\alpha^2 \Omega$ approximations, respectively. The dot-dashed
  ($\alpha \Omega$) and triple-dot-dashed ($\alpha^2 \Omega$) curves
  give the corresponding results with a rotation profile where the
  surface shear layer has been removed. The thick solid line describes
  the Sun, i.e. where $C_\alpha/C_\Omega = 10^{-3}$.}
\label{fig:calpcritsimp}
\end{figure}

Although the main emphasis of this paper is to investigate the role of
anisotropies found in the local convection models, we perform a set of
calculations with a constant $\alpha$-effect in the convection zone
with a previously commonly used $\cos{\theta}$ latitude profile in
order to demonstrate the current problems of solar mean-field dynamo
models. In all of the calculations in this section, we put
\begin{equation}
\alpha_{ij} = \frac{1}{2}\,\alpha_0 \delta_{ij} \bigg[ 1 + \tanh
  \bigg( \frac{r - r_{\rm c}}{d_1} \bigg) \bigg] \cos
\theta\;, \label{equ:simpalpha}
\end{equation}
where $r_{\rm c} = 0.7$ and $d_1 = 0.015$. The $\tanh$-factor takes
care of the vanishing $\alpha$-effect in the overshoot region. We
study this simple case first and determine the critical dynamo number
$C_\alpha^{\rm crit}$ as a function of $C_\Omega$ for the
antisymmetric A0 and symmetric S0 modes. The results for the A0 case
are shown in Fig. \ref{fig:calpcritsimp}. We note that the critical
values for the S0 mode are very close to those of the A0 mode, but
consistently slightly larger. The difference, however, is less than
one per cent in all cases. For values of $C_\Omega$ larger than $10^4$
the difference between the $\alpha \Omega$ and $\alpha^2 \Omega$
models is less than 5 per cent. In what follows we consider the case
$C_\alpha = 15$, $C_\Omega = 1.5 \cdot 10^4$ as our standard model, in
which case we adopt the $\alpha \Omega$-approximation.  Moreover, we
note that the critical dynamo number in the case of no surface shear
is roughly 20 per cent larger than in the case where it is taken into
account.

Keeping the ratio $C_\alpha/C_\Omega$ fixed at $10^{-3}$ we find that
the critical dynamo number, $D^{\rm crit} \equiv C^{\rm crit}_\alpha
C_\Omega$, for the excitation of the antisymmetric A0 mode is roughly
$D^{\rm crit} \approx 4.89 \cdot 10^4$ corresponding to $C_\alpha^{\rm
  crit} \approx 6.99$ using the $\alpha \Omega$-approximation ($D^{\rm
  crit} \approx 4.37 \cdot 10^4$ corresponding to $C_\alpha^{\rm crit}
\approx 6.61$ if the $\alpha$-terms are retained in the equation of
$B$). In the marginal case the turbulent diffusivity turns out to be
$\eta_0 \approx 2.0 \cdot 10^8$\,m$^2$\,s$^{-1}$. The toroidal
magnetic field of the marginal solution exhibits an oscillation period
of $t_{\rm cyc} \approx 0.044\tau$, which corresponds to roughly 3.2
years in physical units.

\begin{figure}
\resizebox{\hsize}{!}
{\includegraphics{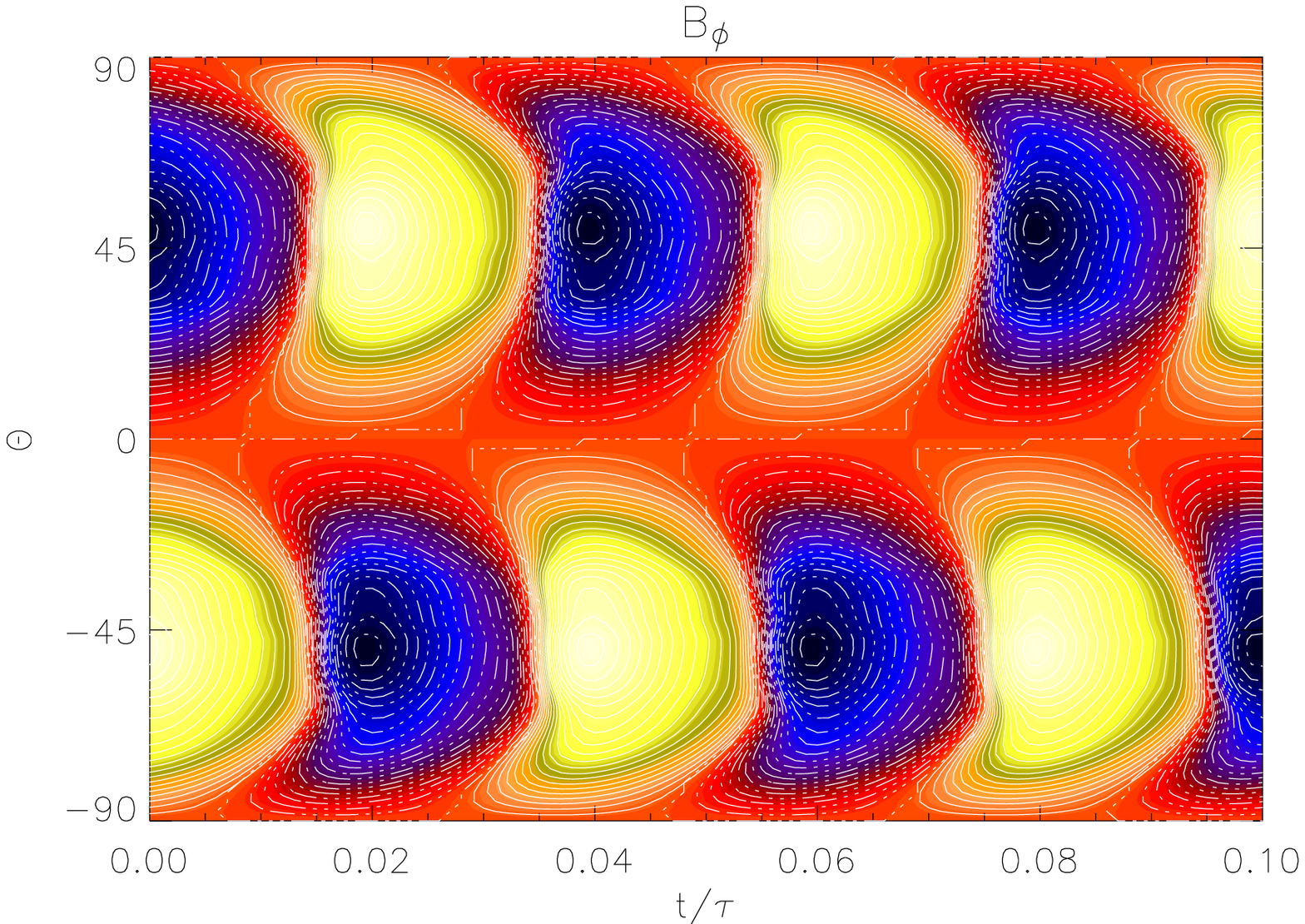}}
\caption{Butterfly diagram of the toroidal magnetic field from $r =
  0.99R$ from a model with $C_\alpha = 15$ and $C_\Omega = 1.5 \cdot
  10^4$ with the rotation profile shown in Fig. \ref{fig:rotproheli}
  with the surface shear at $r > 0.95R$ artificially turned off.}
\label{fig:cosabutter_nss}
\end{figure}

\begin{figure}
\resizebox{\hsize}{!}
{\includegraphics{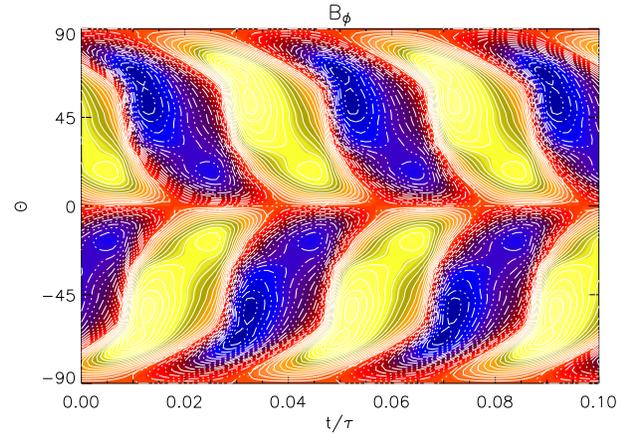}}
\caption{Butterfly diagram of the toroidal magnetic field from $r =
  0.99R$ from a model with $C_\alpha = 15$ and $C_\Omega = 1.5 \cdot
  10^4$ with the full rotation profile shown in
  Fig. \ref{fig:rotproheli}. The cycle period is approximately $t_{\rm
    cyc} \approx 6.5$ years.}
\label{fig:cosabutter}
\end{figure}

Considering the resulting butterfly diagrams we compare models where
the shear above $r = 0.95R$ is either turned off
(Fig.~\ref{fig:cosabutter_nss}) or retained
(Fig.~\ref{fig:cosabutter}). In the former case the butterfly diagram
of the azimuthal field near the surface shows that the migration of
the activity belts at latitudes $|\Theta| < 60\degr$ is poleward with
an equatorward branch near the pole, whereas the latter case exhibits
a strong equatorward propagating activity belt from latitude $\approx
80 \degr$ all the way down to the equator. The migration of the former
type has frequently been reported in mean-field models with
distributed $\alpha$-effect combined with helioseismological rotation
profile (e.g. Bonanno et al. 2002).  Concerning the latter case, we
note here that if there is a non-vanishing $\alpha$-effect near the
surface, the surface shear layer plays a very important role in
shaping the butterfly diagram. However, if the $\alpha$-effect is more
concentrated in the deep layers of the convection zone the surface
shear plays only a minor role (see Sect.~\ref{subsec:surfshear}).

\begin{figure*}
\resizebox{\hsize}{!}
{\includegraphics{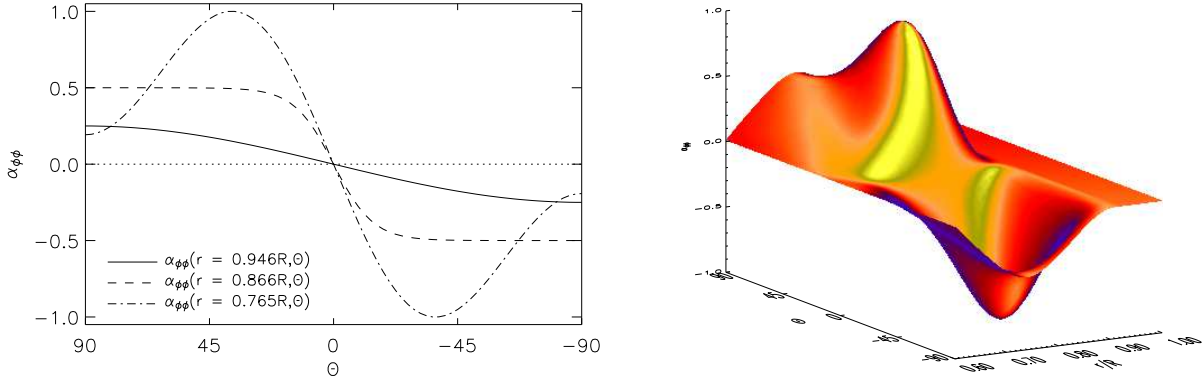}}
\caption{Left panel: profiles of the $\alpha_{\phi \phi}$-component
  from three depths $r = (0.765, 0.866, 0.946)R$ corresponding to
  Coriolis numbers ${\rm Co} = (1,4,10)$, respectively, using the
  convective velocity at the bottom of the convection zone as
  normalisation. Right panel: the actual $\alpha_{\phi
    \phi}(r,\theta)$ profile used in the models.}
\label{fig:convalpp}
\end{figure*}

\subsection{$\alpha$-effect from the local convection models}
Already the early analytical investigations of the turbulent
electromotive force showed that the $\alpha$-effect is highly
anisotropic (Steenbeck, Krause \& R\"adler
\cite{Steenbeckea1966}). These anisotropies have recently been studied
by means of local convection calculations by Ossendrijver et
al. (\cite{Osseea2002}) and KKOS. In the present study we employ these
numerical results of the $\alpha$-effect and turbulent pumping in
solar dynamo models. As was discussed earlier, we set the ratio
$C_\alpha/C_\Omega$ to $10^{-3}$ in the case of the solar
dynamo. Effectively the small value of the ratio $C_\alpha/C_\Omega$
means that the dominant (diagonal) component of
$\overline{\vec{\alpha}}$ is $\alpha_{\phi \phi}$ which appears in the
equation of the poloidal field (see Eqs.~\ref{equ:A} and
\ref{equ:B}). Adopting the $\alpha \Omega$-approximation and setting
$\alpha_{\rm rr} = \alpha_{\theta \theta} = 0$ does not significantly
change our results. These components can, however, be of vital
importance for $\alpha^2$-dynamos and their oscillation properties
such as those investigated by R\"udiger, Elstner \& Ossendrijver
(\cite{Ruedigerea2003}).

In order to make use of the local results of KKOS we use the
Coriolis number to determine the depth of the local convection model
in the convection zone. In an earlier study we computed the Coriolis
number from a mixing-length model of the solar convection zone (Fig.~1
of K\"apyl\"a et al. \cite{Kaepylaeea2005}), and found that it varies
between $10^{-3}$ near the surface and $10$ or larger in the deep
layers. In KKOS, sets of convection calculations at different
latitudes are made with approximate Coriolis numbers of 1, 4, and
10. Identifying these values with depths in the solar convection zone,
they correspond to $r = 0.946R$, $r = 0.866R$, and $r = 0.765R$,
respectively. In KKOS it was found that for Co = 1 the component
corresponding to $\alpha_{\phi \phi}$ is consistent with a $\cos
\theta$ latitude profile, whereas for Co = 4 $\alpha_{\phi \phi}$ is
basically constant as function of latitude from the poles to the
latitude $|\Theta| = 30 \degr$. In the rapid rotation regime, i.e. Co
= 10, $\alpha_{\phi \phi}$ peaks around $|\Theta| = 30 \degr$ and the
magnitude at the poles decreases. We use the following approximations
of the latitude dependence of $\alpha_{\phi \phi}$ at the different
depths
\begin{eqnarray}
\alpha_{\phi \phi}(r = 0.946R,\theta) & = & \frac{1}{4} \alpha_0 \cos \theta\;, \label{equ:alpp0946} \\ 
\alpha_{\phi \phi}(r = 0.866R,\theta) & = & -\frac{1}{2}\alpha_0 \tanh \bigg[ 4 \Big(\theta - \frac{\pi}{2} \Big) \bigg]\;, \label{equ:alpp0866} \\ 
\alpha_{\phi \phi}(r = 0.765R,\theta) & = & 3\,\alpha_0 \Big(\sin^2 \theta \cos \theta+ \frac{1}{4} \cos \theta \Big)\;, \label{equ:alpp0765}
\end{eqnarray}
see the left panel of Fig.~\ref{fig:convalpp} (see also Figs.~3 to 7
of KKOS). We use cubic spline interpolation in order to obtain a value
of $\alpha_{\phi \phi}$ for all grid points, see the right panel of
Fig.~\ref{fig:convalpp}. In Eqs.~(\ref{equ:alpp0946}) to
(\ref{equ:alpp0765}) an estimate of the convective velocity, $u
\approx 10$\,m\,s$^{-1}$, at the bottom of the convection zone has
been used to scale the profiles at all depths resulting an
$\alpha_{\phi \phi}$ profile that peaks near the bottom of the
convection zone at around latitude $30\degr$.

\begin{figure}
\resizebox{\hsize}{!}
{\includegraphics{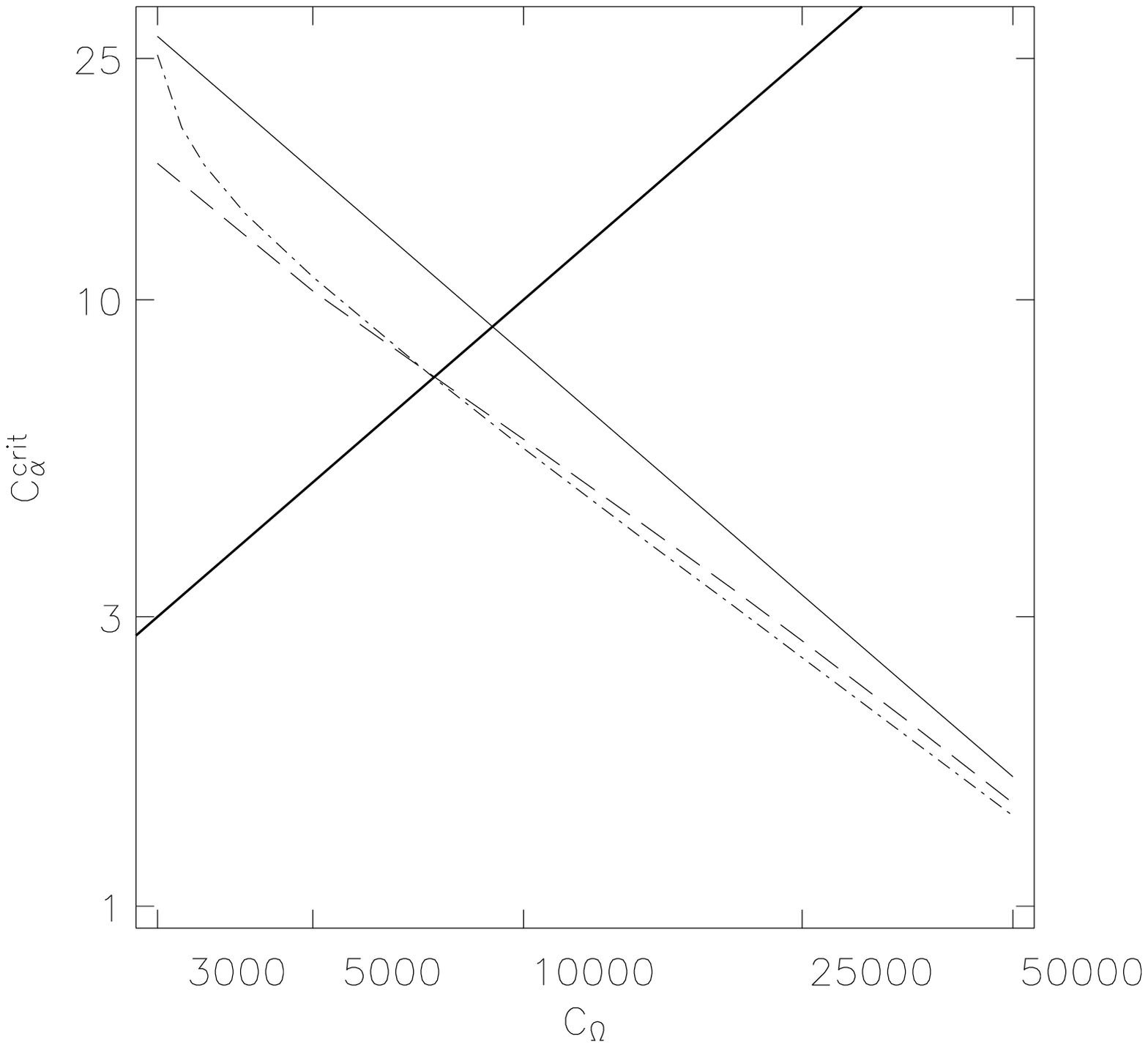}}
\caption{Critical dynamo number $C_\alpha^{\rm crit}$ as a function of
  $C_\Omega$ for the antisymmetric A0 mode with the $\alpha$-profile
  from convection calculations, see Fig. \ref{fig:convalpp} (solid
  line). The additional curves are with latitudinal and radial pumping
  according to Eqs.~(\ref{equ:gammar}) and (\ref{equ:gammat}) (dashed
  line), and with meridional flow according to Eqs.~(\ref{equ:ur}) and
  (\ref{equ:ut}) (dot-dashed line). The thick solid line describes the
  Sun, i.e. where $C_\alpha/C_\Omega = 10^{-3}$.}
\label{fig:calpcritconv}
\end{figure}

Considering models with the $\alpha$-effect depicted in
Fig.~\ref{fig:convalpp}, we find that the critical dynamo number is
significantly larger than in the case of constant $\alpha$ in the
convection zone, see Fig.~\ref{fig:calpcritconv}. Again the critical
value for the S0 mode is very sligthly larger than that of the A0
mode. For the solar case of $C_\alpha/C_\Omega = 10^{-3}$ we find
$D^{\rm crit} \approx 8.2 \cdot 10^4$, or $C_\alpha^{\rm crit} =
9.04$, instead of $D^{\rm crit} \approx 4.89 \cdot 10^4$ and
$C_\alpha^{\rm crit} \approx 6.99$ in the case of constant
$\alpha_{\phi \phi}$ in the whole convection zone. This is of course
due to the fact that the integral of $\alpha_{\phi \phi}$ over the
full shell is significantly less than in the case of radially constant
$\alpha_{\phi \phi}$. Furthermore, we find that initially purely
antisymmetric A0-type configurations are only stable when $C_\alpha <
12$. In the range $12 < C_\alpha < 17$ the initially purely dipolar A0
solutions evolve towards a symmetric S0 solution after about one
diffusion time. Furthermore, when $C_\alpha > 17$ mixed mode solutions
are excited. The butterfly diagrams for the cases $C_\alpha < 17$ are
characterised by strong activity at high latitudes, the maximum being
near $50\degr$, see Fig.~\ref{fig:convbutter}. The migration of the
activity belts is equatorward at high latitudes and poleward from the
equator up to latitude of about $45\degr$. This demonstrates that the
included surface shear layer, leading to equatorward migration in the
case of constant in radius $\alpha$-effect, plays a less important
role when $\alpha$ concentrates at the bottom of the convection
zone. The parameter range within which the dipolar A0 solution is
stable in the nonlinear regime is quite narrow. Adding radial and
latitudinal components of the pumping vector or a meridional flow
pattern is observed to widen this range (see below).

\subsubsection{Effect of the general pumping effect}
The local convection models of KKOS yield all the three components of
the pumping vector $\vec{\gamma}$. In the present axisymmetric
investigation, the azimuthal pumping $\gamma_\phi$, however, would
appear only through its radial and latitudinal gradients. These terms
would act analogously to the differential rotation but due to the
small value of $C_\alpha/C_\Omega$, these gradients are negligible in
comparison to the differential rotation. Therefore we consider here
only the radial and latitudinal pumping effects.

According to the local models, the radial pumping is directed
downwards throughout the convection zone in the slow and moderate
rotation regimes, i.e. ${\rm Co} < 4$. The radial pumping velocity is
non-zero also in the case of no rotation due to the fact that it is
mostly caused by the diamagnetic effect (see e.g. Ossendrijver et
al. \cite{Osseea2002}). In the slow and moderate rotation regime the
dependence on rotation (depth) is indeed rather weak, although
$\gamma_r$ tends to somewhat increase at the poles and decrease at the
equator as function of rotation. The trend is most clear in the rapid
rotation regime, i.e. Co = 10, where the radial pumping is very small
or even positive at low latitudes and at the equator. In the overshoot
region $\gamma_r$ vanishes. We adopt a simplified profile which
captures the main aspects of the results of KKOS by setting
\begin{eqnarray}
\gamma_r(r,\theta) & = & -\frac{1}{2} \alpha_0
\bigg[\tanh{\Big(\frac{r - r_c}{d_1}\Big)} - \tanh{\Big(\frac{r -
      r_1}{d_2}\Big)}\bigg] \times \nonumber \\ &&
\bigg\{\exp{\bigg[\frac{(r - r_c)^2}{d_3^2}}\bigg]
\sin{\theta} + 1\bigg\}\;, \label{equ:gammar}
\end{eqnarray}
where $r_c=0.7$, $r_1 = 0.975$, $d_1 = 0.015$, $d_2 = 0.1$, and $d_3 =
0.25$. The maximum magnitude of the vertical pumping is taken to be
the same as that of the $\alpha$-effect, i.e. $\alpha_0$, as is
suggested by the local results..

\begin{figure}
\resizebox{\hsize}{!}
{\includegraphics{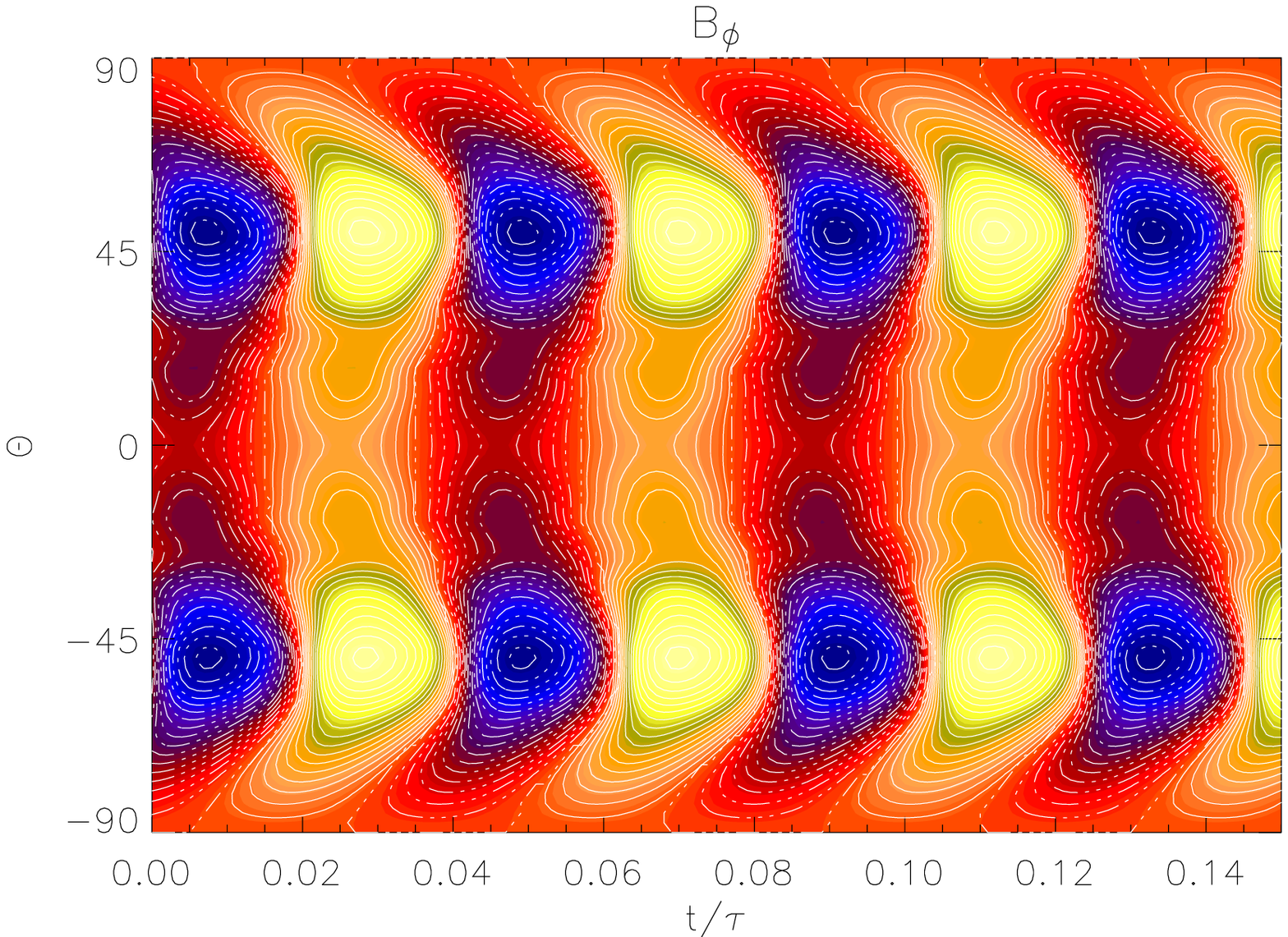}}
\caption{Butterfly diagram of the toroidal magnetic field from $r =
  0.99R$ from a model with $C_\alpha = 15$ and $C_\Omega = 1.5 \cdot
  10^4$. The cycle period is approximately $t_{\rm cyc} \approx 7.0$
  years. $\alpha_{\phi \phi}$-profile as shown in
  Fig.~\ref{fig:convalpp}.}
\label{fig:convbutter}
\end{figure}

The latitudinal pumping velocity $\gamma_\theta$ is directed
equatorwards throughout the convection zone. This effect is
rotationally excited (Krause \& R\"adler \cite{KrauseRaedler1980}),
and thus it shows a strong rotational dependence so that in the bottom
of the convection zone it has a peak magnitude of $\approx
2$\,m\,s$^{-1}$ whilst it is negligible at the surface. Latitudinally
$\gamma_\theta$ is strongest near the equator at around $15\degr$ with
decreasing values towards the poles. Similarly to the radial
component, $\gamma_\theta$ vanishes in the overshoot region. We adopt
again a simplified profile described by
\begin{eqnarray}
\gamma_{\theta}(r,\theta) & = & \alpha_0 \bigg[\tanh{\Big(\frac{r -
      r_c}{d_1}\Big)} - \tanh{\Big(\frac{r - r_2}{d_4}\Big)}\bigg]
\times \nonumber \\ &&
\sin^4{\theta}\cos{\theta}\;, \label{equ:gammat}
\end{eqnarray}
where $r_2 = 0.875$ and $d_4 = 0.075$. The magnitude of the pumping
effect in the mean-field equations is controlled by the nondimensional
parameter $C_\gamma$; since the pumping effect is basically a part of
the $\alpha$-effect and the maximum magnitudes of both effects were
very similar in the study of KKOS we set $C_\gamma = C_\alpha$.

\begin{figure}
\resizebox{\hsize}{!}
{\includegraphics{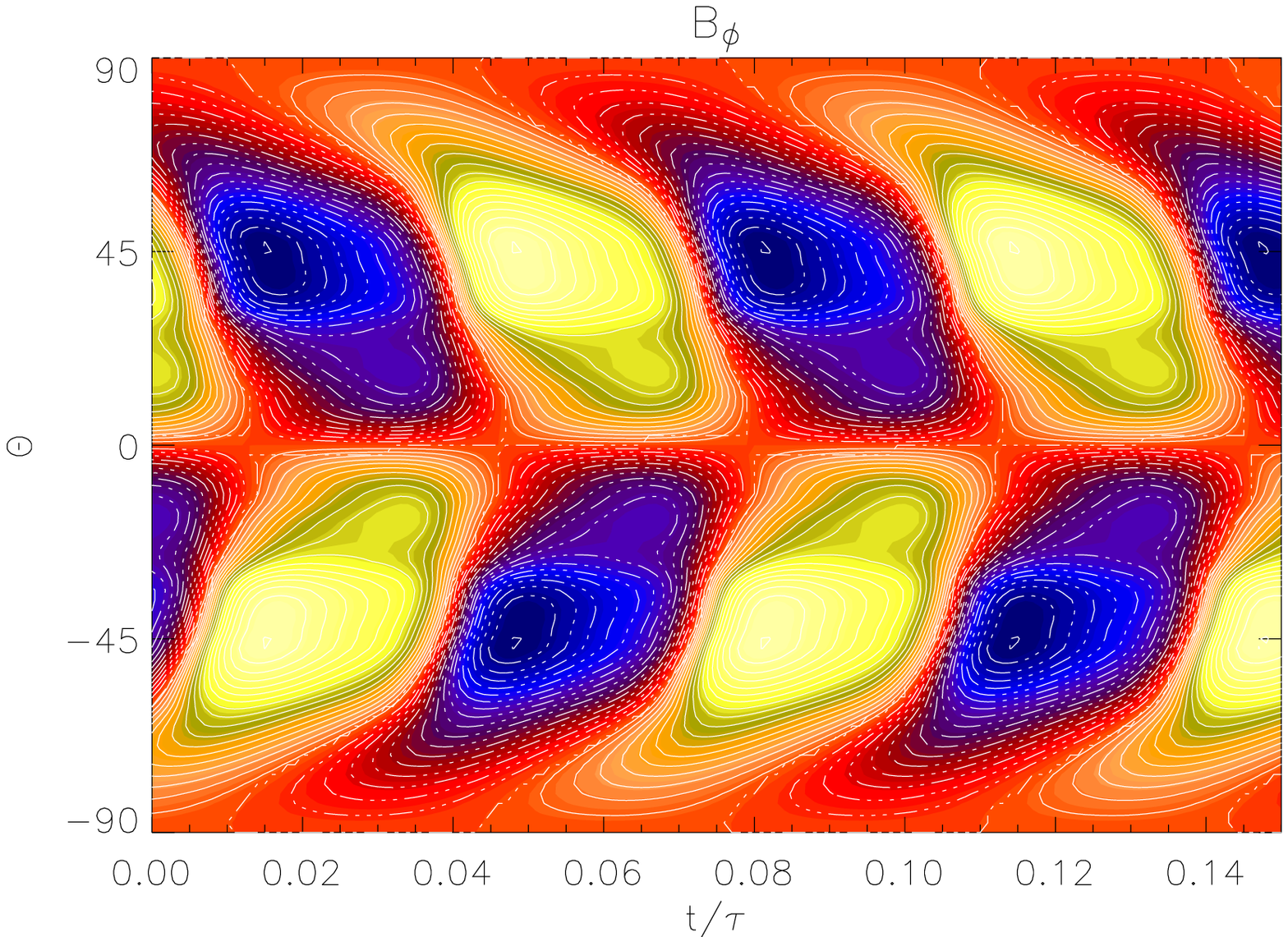}}
\caption{Same as Fig.~\ref{fig:convbutter} but with added radial and
  latitudinal pumping velocities as defined via
  Eqs.~(\ref{equ:gammar}) and (\ref{equ:gammat}). The cycle period is
  $t_{\rm cyc} \approx 12$ years.}
\label{fig:convbutterpump}
\end{figure}

Adding the pumping velocities defined via Eqs.~(\ref{equ:gammar}) and
(\ref{equ:gammat}) into the model already employing the
$\alpha$-effect from convection calculations, widens the region in
which the dipolar solution is the preferred one (up to $C_\alpha
\approx 50$). The dashed line in Fig.~\ref{fig:calpcritconv} indicates
that the pumping effects make the dynamo easier to excite. This is
logical, due to the fact that the radial pumping is generally
downward, i.e. towards regions of larger $\alpha$-effect. The
latitudinal pumping in the deeper layers has a similar effect, i.e. it
transport magnetic fields to lower latitudes. Furthermore, using our
standard values $C_\alpha = 15$ and $C_\Omega = 1.5 \cdot 10^4$, the
cycle period is also significantly longer in this case, i.e. $t_{\rm
  cyc} \approx 11$ years instead of approximately seven years in the
case without the pumping effects. Another clear effect is that the
migration of the activity belts is now equatorward also at low
latitudes, see Fig.~\ref{fig:convbutterpump}. Both of these results
seem to be direct consequences of the added latitudinal pumping;
otherwise identical model with $\gamma_\theta = 0$ yields a butterfly
diagram very similar to Fig.~\ref{fig:convbutter}.

\subsubsection{Effect of field direction dependent pumping effect}
\label{sec:pumpfdi}
\begin{figure}
\resizebox{\hsize}{!}
{\includegraphics{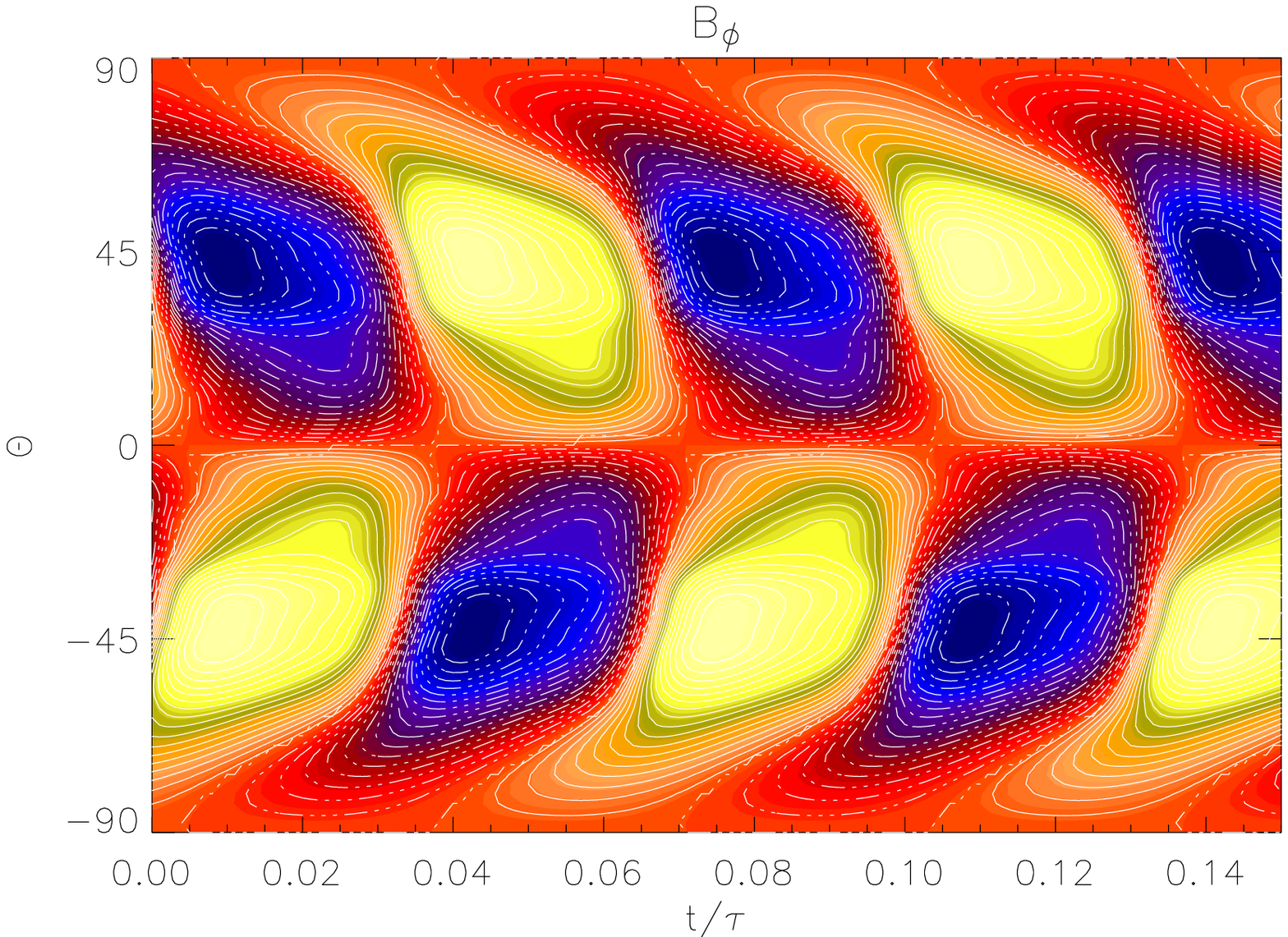}}
\caption{Same as Fig.~\ref{fig:convbutterpump} but with added field
  direction dependent pumping according to
  Eq.~(\ref{equ:gammafdi}). The cycle period is about 12 years.}
\label{fig:convbutterpumpfdi}
\end{figure}

The off-diagonal components of the (symmetric) tensor
$\overline{\vec{\alpha}}$ contribute to the field direction dependent
pumping (Kitchatinov \cite{Kitchatinov1991}; Ossendrijver et
al. \cite{Osseea2002}; KKOS). Therefore, in the general case, we need
to introduce $\alpha_{ij}$ components $\alpha_{r \theta}$, $\alpha_{r
  \phi}$, and $\alpha_{\phi \theta}$ into the model. However,
according to the results of KKOS, $\alpha_{\phi \theta}$ is clearly
the largest in magnitude of the components entering the equation for
$A$. The local calculations show that this component peaks strongly
near the equator at or around latitude $15\degr$, and the maximum
absolute magnitude is again roughly 2\,m\,s$^{-1}$. We introduce a
profile
\begin{eqnarray}
\alpha_{\phi \theta}(r,\theta) & = & -\alpha_0 \bigg[\tanh{\Big(\frac{r
      - r_c}{d_1}\Big)} - \tanh{\Big(\frac{r - r_3}{d_2}\Big)}\bigg]
\times \nonumber \\ && \sin^{12} \theta \cos^2 \theta\;, \label{equ:gammafdi}
\end{eqnarray}
where $r_3 = 0.85$ and $d_2 = 0.1$. The main effect of $\alpha_{\phi
  \theta}$ is that it contributes to the radial pumping of the
azimuthal field via $\gamma_r^{\phi} = \gamma_r - \alpha_{\phi
  \theta}$, where $\gamma_r^{\phi}$ is the pumping velocity for the
azimuthal field. The local results of KKOS indicate that the azimuthal
field can be pumped \emph{upward} at latitudes $|\Theta| <
45\degr$. Due to the narrow latitude and radius range where
$\alpha_{\phi \theta}$ is appreciably large, its effects on the
surface are hardly visible (compare Figures \ref{fig:convbutterpump}
and \ref{fig:convbutterpumpfdi}).

\subsection{Effect of meridional flow}
Solar surface indicators and local helioseismology (Zhao \& Kosovichev
\cite{ZhaoKosov2004}; Komm et al. \cite{Kommea2004}) have revealed
that there is a mean poleward flow of 10--20\,m\,s$^{-1}$ near the
solar surface. Although no observational evidence exists as of yet, it
is generally accepted that there is a return flow at some greater
depth. However, it is also not known whether there is only one cell in
radius or more. The simplest case is to assume that there is only one
cell. We adopt a modified version of the flow used by Dudley \& James
(\cite{DudleyJames1989}), defined via a potential for the poloidal
flow
\begin{eqnarray}
\psi = x^2 \sin(\pi x)P_2^0 (\cos \theta)\;,
\end{eqnarray}
where $x = (r - r_c)/(R - r_c)$, and $P_2^0(\cos \theta) =
\frac{1}{2}(3\cos^2 \theta - 1)$. Here $x$ describes the effect of
decreasing density towards the top of the convection zone. The radial
and latitudinal velocity components are obtained via
\begin{eqnarray}
u_r & = & -\frac{1}{r^2 \sin \theta} \frac{\partial}{\partial \theta} \Big( \sin \theta \frac{\partial \psi}{\partial \theta} \Big)\;, \\
u_\theta & = & \frac{1}{r} \frac{\partial}{\partial r} \frac{\partial \psi}{\partial \theta}\;,
\end{eqnarray}
which are written out explicitly in Eqs.~(\ref{equ:ur}) and
(\ref{equ:ut}) in Appendix A.

\begin{figure}
\resizebox{\hsize}{!}
{\includegraphics{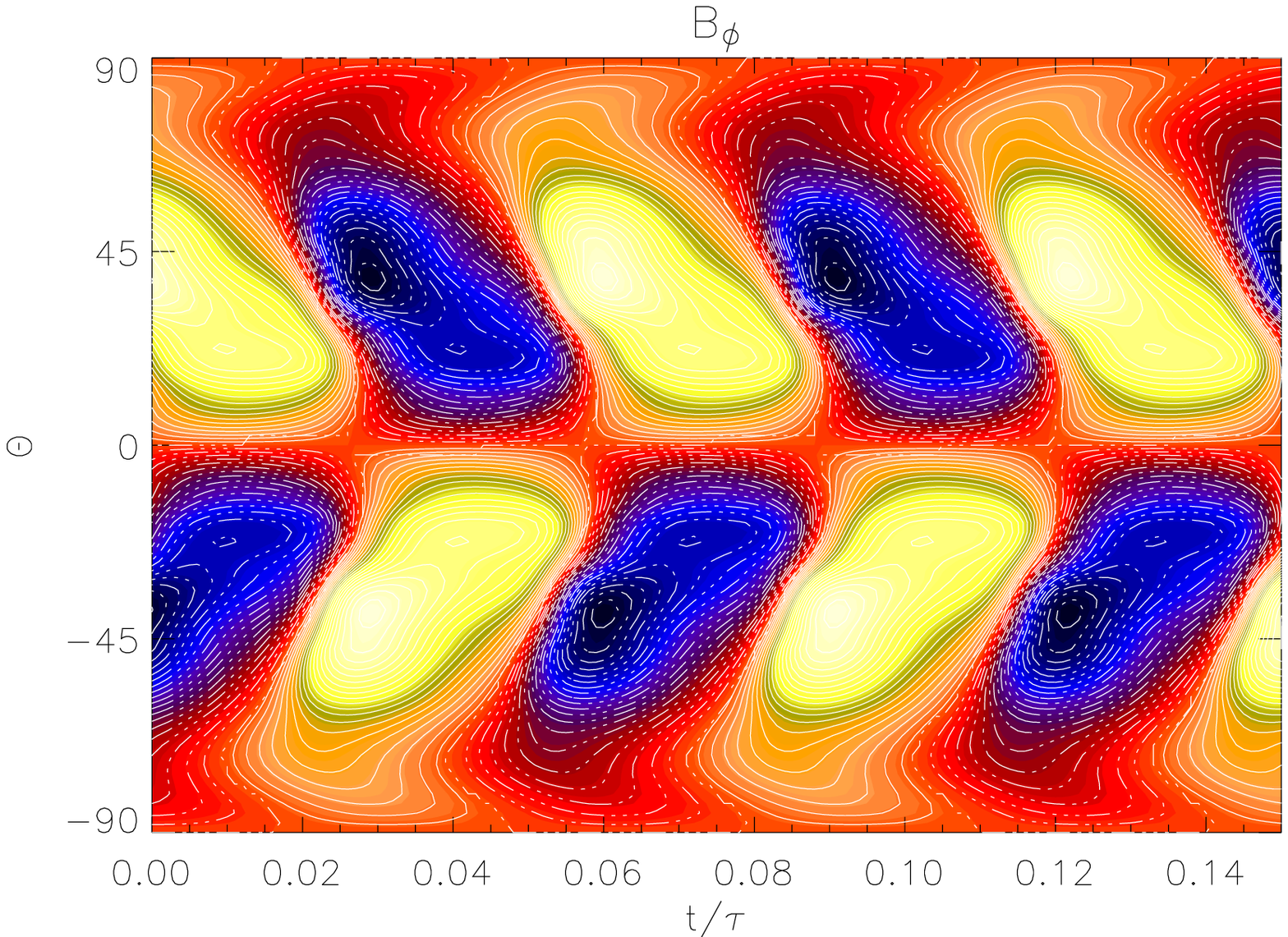}}
\caption{Same as Fig.~\ref{fig:convbutterpumpfdi} but with added
  meridional flow with max($u_\theta$) = 10\,m\,s$^{-1}$ at the
  surface, with which choice $C_U = 75$. The cycle period is about
  11.5 years.}
\label{fig:convbutterpumpfdimeri}
\end{figure}

Fixing the maximum value of the flow at the surface to 10\,m\,s$^{-1}$
yields $C_U = 75$ or $C_U/C_\Omega = 0.005$ for the standard model
where $C_\alpha = 15$. Introducing this flow into the model discussed
in Sect.~\ref{sec:pumpfdi} which already includes both pumping
effects, we obtain the butterfly diagram shown in
Fig.~\ref{fig:convbutterpumpfdimeri}. The most visible effect of the
meridional flow is the appearance of the polar branch at latitudes
$\Theta > 60\degr$. Furthermore, the equatorward propagating activity
belts appear now at slightly lower latitudes, although the difference
to the observed solar activity is still quite large. The critical
dynamo numbers in the case with meridional flow are in general
somewhat higher than in the case with only radial and latitudinal
pumping taken into account. Furthermore, when $C_\Omega$ is smaller
than approximately $1.5 \cdot 10^{4}$, the symmetric S0 mode is
somewhat easier to excite than the A0 mode (not shown).

In the nonlinear regime the solution is dipolar (A0) for $C_\alpha <
15$, quadrupolar in the range $C_\alpha \approx 15 \ldots 30$, and
mixed parity solutions are obtained when $C_\alpha > 30$ keeping the
ratio $C_U/C_\Omega = 0.005$ fixed. If $C_U/C_\Omega$ is incereased
the activity belts are shifted closer to the equator, but the S0 and
mixed modes become somewhat easier to excite. Increasing the dynamo
numbers is associated with an increasing cycle period, but with the
present profiles for the $\alpha$-effect and turbulent pumping the
solutions in the solar range, i.e. $t_{\rm cyc} \approx 22$ years, all
consist of mixed modes.

We note here that in the present models the magnetic Reynolds number
is equal to $C_U$, i.e. ${\rm Rm} = C_U \approx
\mathcal{O}(100)$. This means that we are not in the advection
dominated regime realised in many flux transport dynamo models
(e.g. Dikpati \& Charbonneau \cite{DikChar1999}; Bonanno et
al. \cite{Bonannoea2002}) where values in excess of $10^3$ are
common. The value of Rm, however, indicates that the present models
are also not fully diffusion dominated, but rather represent a
distributed dynamo.

\subsection{The effect of the surface shear layer}
\label{subsec:surfshear}
\begin{figure}
\resizebox{\hsize}{!}
{\includegraphics{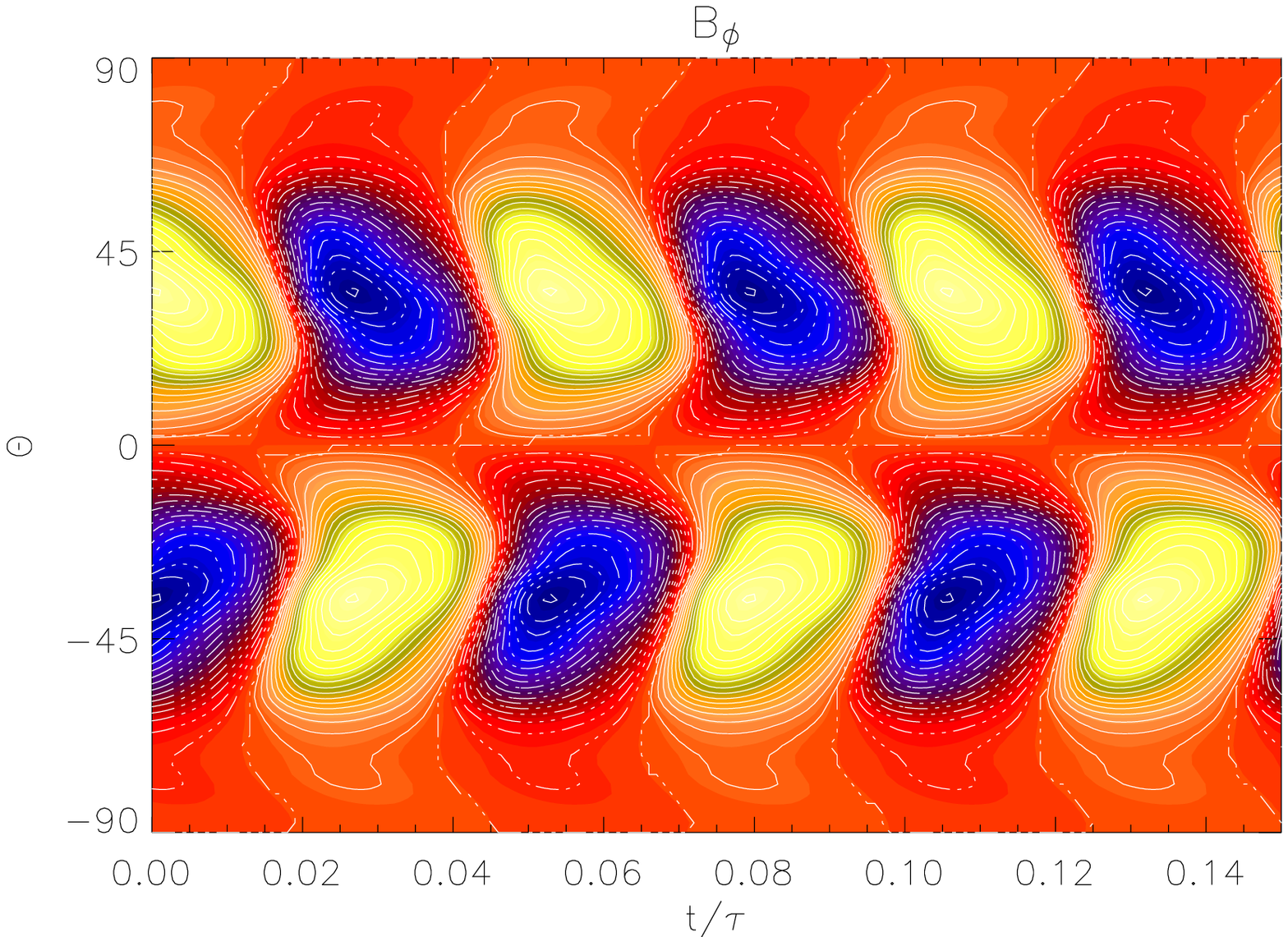}}
\caption{Same as Fig.~\ref{fig:convbutterpumpfdimeri} where the
  surface shear above $r = 0.95R$ has been artificially turned
  off. The cycle period is about 9.5 years.}
\label{fig:convbutter_nss}
\end{figure}

In order to study the influence of the surface shear layer on the
solution we take the same setup as in the previous section and turn
off the shear above $r = 0.95R$. The resulting butterfly diagram is
shown in Fig.~\ref{fig:convbutter_nss}. Comparing this with
Fig.~\ref{fig:convbutterpumpfdimeri} shows some differences, most
visibly the less extended active regions, and a narrow region of
poleward migration near the equator. In contrast to the case of
uniformly distributed $\alpha$-effect (see
Sect.~\ref{subsec:simpalpha}) for which case the surface shear plays a
significant role, the importance of the surface shear is very small if
the $\alpha$-effect more concentrated near the bottom of the
convection zone.

\subsection{Phase relation of $B_r B_\phi$}
\begin{figure}
\resizebox{\hsize}{!}
{\includegraphics{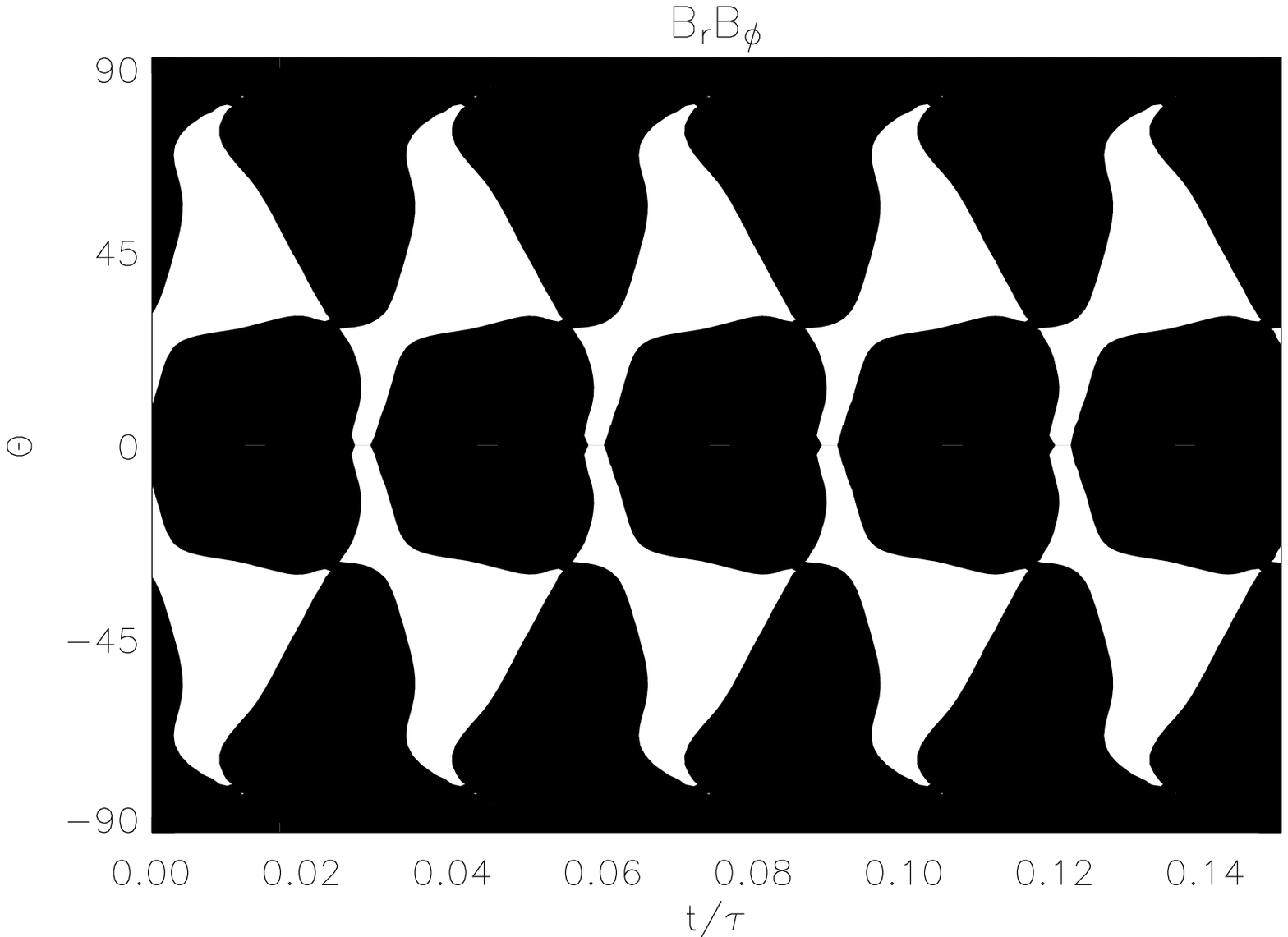}}
\caption{The correlation $B_r B_\phi$ from the model shown in
  Fig.~\ref{fig:convbutterpumpfdimeri}. Black denotes negative and
  white positive values.}
\label{fig:phase}
\end{figure}

We note briefly that in our models the correlation of the radial and
azimuthal magnetic field components, i.e. $B_r B_\phi$, is mostly
negative (see Fig.~\ref{fig:phase} for a typical result) in the
latitude range where the activity belts are located in accordance with
solar observations (e.g. Stix \cite{Stix1976}). This relation has
commonly been used as a constraint for solar dynamo models (e.g. Stix
\cite{Stix1976}; Schlichenmaier \& Stix \cite{SchlicheStix1995};
Brandenburg \cite{Brandenburg2005}) although this has recently been
criticized by Sch\"ussler (\cite{Schuessler2005}) who explains this
relation as the consequence of the systematic tilt of the bipolar
regions due to the action of Coriolis force on rising flux ropes in
the convection zone.

\section{Conclusions}
\label{sec:conclusions}
We have studied kinematic mean-field solar dynamo models using
helioseismologically determined rotation profile and $\alpha$-effect
and turbulent pumping from the local convection calculations of
K\"apyl\"a et al. (\cite{Kaepylaeea2006}). We show that if a simple
$\alpha$-effect profile proportional to $\cos \theta$ in latitude and
constant in radius is applied, the results are very sensitive to the
presence of a surface shear layer at $r > 0.95R$. Omitting this
feature, the migration of the activity belts is equatorward near the
poles, weak at midlatitudes, and poleward at low latitudes. The
inclusion of the shear layer results in equatorward propagation at
virtually all latitudes.

The $\alpha$-coefficients from local convection calculations tend to
be much more concentrated near the bottom of the convection zone and
also closer to the equator with peak values occuring at around
latitude $30\degr$. In this case, however, the critical dynamo numbers
are significantly larger due to the more concentrated nature of the
$\alpha$-effect. The solution is dipolar when $C_\alpha < 12$ and
quadrupolar (S0) if $C_\alpha$ is larger even if a purely dipolar
initial condition is used. Adding a latitudinal pumping velocity of
the order of 2\,m\,s$^{-1}$ at the bottom of the convection zone
remedies this problem and helps to shift the activity belts closer to
the equator, whereas the vertical pumping seems to have a much smaller
overall signifigance for the stability and overall structure of the
solution. The latitudinal pumping also helps to restore the
equatorward migration of activity belts, lost by the implementation of
the $\alpha$-effect from convection calculations. Adding the pumping
effects also makes the cycle period significantly longer. Additional
upward pumping of the azimuthal field component through the component
$\alpha_{\phi \theta}$ at near equator regions has very little effect
on the solution. In all the models investigated, the correlation $B_r
B_{\phi}$ is mostly negative, which is in agreement with the
observations, but none of the models produces poleward migration at
high latitudes (polar branch).

If a single cell meridional flow patterns where the flow is of the
order of 10\,m\,s$^{-1}$ and poleward near the surface and
significantly smaller and equatorward at larger depths helps to shift
the activity belts further closer to the equator, although they still
appear too high ($|\Theta| \approx 5\degr \ldots 60\degr$) in
comparison to the Sun. A region of poleward migration also appears at
high latitudes.  If the upper shear layer is removed from this model,
the butterfly diagram changes only little in contrast to the models
where there was an appreciably large $\alpha$-effect also near the
surface.

The distributed dynamo model investigated in this paper correctly
reproduces many of the general features of the solar cyclic activity,
including realistic migration patterns and correct phase relation, but
some problems persist. These include the excitation of the dipolar
(A0) solutions only within a narrow range of parameters with the
anisotropic $\alpha$-effect combined with meridional circulation,
activity at too high latitudes for all models investigated, and the
somewhat too short dynamo cycles. With further tuning of the various
parameters going into the model (magnitudes of $\alpha$- and
$\gamma$-effects, strength and profile of the meridional circulation)
it would most likely be possible to obtain a solution reproducing the
missing features, as well. Instead of going further into the domain of
tuning in the present study, we note that further work in the context
of local convection modelling is needed in order to obtain the totally
missing information on the anisotropies of turbulent diffusivity (work
in progress). It is also necessary to consider the nonlinear
saturation process of the $\alpha$-effect in more detail, for instance
in the form of a dynamical $\alpha$-effect with helicity fluxes
(e.g. Brandenburg \& Subramanian \cite{BranSubr2005}). Further
improvements of the present mean-field model include the relaxation of
the kinematic approach to include full dynamics, for which purpose
information from local convection calculations also exist
(e.g. K\"apyl\"a, Korpi \& Tuominen \cite{Kaepylaeea2004}). With the
inclusion of hydro- and thermodynamics the nonlinear feedbacks of the
kind investigated by e.g. Rempel (\cite{Rempel2006}) could be included
in a self-consistent way.

Despite of the shortcomings discussed above, we feel that the simple
model presented in this paper demonstrates two important aspects
concerning solar dynamo theory. Firstly, as recently discussed at
length by Brandenburg (\cite{Brandenburg2005}) the surface shear layer
may be important in shaping the solar dynamo; according to our results
this is indeed to be expected if the $\alpha$-effect has an
appreciable magnitude near the surface. Secondly, we are able to
demonstrate that a distributed $\alpha \Omega$-dynamo model applying a
realistic rotation profile and meridional flow can reproduce the
correct phase relation and migration of the activity belts. Although
meridional circulation is an essential ingredient also in our model,
the magnitude of this effect is of the same order as the turbulent
inductive effects and the turbulent diffusivity is an order of
magnitude larger than in the advection dominated flux-transport
dynamos.

\acknowledgements{PJK acknowledges the financial support from the
Finnish graduate school for astronomy and space physics and the
Kiepenheuer-Institut and the Academy of Finland grant No. 1112020 
for travel support.}

\begin{appendix}
\section{Equations}
The equation for $A$ reads
\begin{eqnarray}
\frac{\pd A}{\pd \tau} & = &  (C_{\alpha} \alpha_{\phi \theta} - C_U U_r) \bigg(\frac{A}{r} + \frac{\pd A}{\pd r} \bigg) \nonumber \\
&+&  (C_{\alpha} \alpha_{\phi r} - C_U U_\theta) \bigg(\frac{\cot \theta}{r} A  + \frac{1}{r}\frac{\pd A}{\pd \theta} \bigg)  \nonumber \\ 
&+& C_{\alpha}\alpha_{\phi \phi}B \nonumber \\ 
&+& \eta_t \bigg(\frac{\pd^2A}{\pd r^2} + \frac{2}{r}\frac{\pd A}{\pd r} + \frac{1}{r^2}\frac{\pd^2 A}{\pd \theta^2} + \frac{\cot \theta}{r^2}\frac{\pd A}{\pd \theta} \nonumber \\ 
&-& \frac{A}{r^2 \sin^2 \theta} \bigg)\;, \label{equ:A}
\end{eqnarray}
where $U_i = U^{mer}_i + \frac{C_{\gamma}}{C_U}\gamma_i$.
\begin{eqnarray}
\frac{\pd B}{\pd \tau}
    & = & C_{\Omega} r \sin \theta \bigg[ \frac{\pd \Omega}{\pd r} \Big(\frac{\cot \theta}{r} A + \frac{1}{r} \frac{\pd A}{\pd \theta} \Big) \bigg] \nonumber \\
    & - & C_{\Omega} \sin \theta \bigg[ \frac{\pd \Omega}{\pd \theta} \Big( \frac{A}{r} + \frac{\pd A}{\pd r} \Big) \bigg] \nonumber \\
    & - & C_U \frac{U_r B}{r} + C_U \frac{1}{r} \frac{\pd U_\phi}{\pd r} \Big(\cot \theta A + \frac{\pd A}{\pd \theta} \Big)  \nonumber \\
    & + & C_U \frac{U_\phi}{r} \Big( \cot \theta \frac{\pd A}{\pd r} - \frac{1}{r} \frac{\pd A}{\pd \theta} \Big)
      - C_U \frac{1}{r} \frac{\pd U_\phi}{\pd \theta} \Big( \frac{A}{r} + \frac{\pd A}{\pd r} \Big)   \nonumber \\
    & - & C_U \bigg( \frac{\pd U_r}{\pd r} B + U_r \frac{\pd B}{\pd r} \bigg) 
      - C_U \frac{1}{r} \bigg( \frac{\pd U_\theta}{\pd \theta} B + U_\theta \frac{\pd B}{\pd \theta} \bigg) \nonumber \\
    & + & C_{\alpha} \bigg( \frac{\alpha_{\theta r}}{r} + \frac{\pd \alpha_{\theta r}}{\pd r} - \frac{1}{r} \frac{\pd \alpha_{rr}}{\pd \theta} \bigg) \bigg(\frac{\cot \theta A}{r} + \frac{1}{r} \frac{\pd A}{\pd \theta} \bigg) \nonumber \\
    & - & C_{\alpha} \bigg( \frac{\alpha_{\theta \theta}}{r} + \frac{\pd \alpha_{\theta \theta}}{\pd r} - \frac{1}{r} \frac{\pd \alpha_{r \theta}}{\pd \theta} \bigg) \bigg(\frac{A}{r} + \frac{\pd A}{\pd r} \bigg) \nonumber \\
    & + & C_{\alpha} \bigg( \frac{\alpha_{\theta \phi}}{r} + \frac{\pd \alpha_{\theta \phi}}{\pd r} - \frac{1}{r} \frac{\pd \alpha_{r \phi}}{\pd \theta} \bigg)B \nonumber \\
    & - & C_{\alpha} \frac{\alpha_{rr}}{r} \bigg(\frac{A}{r \sin^2 \theta} - \cot \theta \frac{\pd A}{\pd \theta} -  \frac{\pd^2A}{\pd \theta^2} \bigg) \nonumber \\ 
    & - & C_{\alpha} \alpha_{\theta \theta} \bigg(\frac{A}{r^2} - \frac{1}{r} \frac{\pd A}{\pd r} - \frac{\pd^2A}{\pd r^2} \bigg) \nonumber \\
    & - & C_{\alpha} \bigg[ \alpha_{\theta r} \frac{\cot \theta}{r} \bigg(\frac{A}{r} + \frac{\pd A}{\pd r} \bigg) - \alpha_{\theta \phi} \frac{\pd B}{\pd r} + \frac{\alpha_{r \phi}}{r} \frac{\pd B}{\pd \theta} \bigg] \nonumber \\
    & + & \eta_t \bigg(\frac{\pd^2B}{\pd r^2} + \frac{2}{r}\frac{\pd B}{\pd r} + \frac{1}{r^2}\frac{\pd^2B}{\pd \theta^2} + \frac{\cot \theta}{r^2}\frac{\pd B}{\pd \theta} \nonumber \\ 
    & - & \frac{B}{r^2 \sin^2 \theta} \bigg)\;. \label{equ:B}
\end{eqnarray}
The meridional flow components are given by
\begin{eqnarray}
u_r & = & 6 \frac{x^2}{r^2} \sin (\pi x) \bigg( \frac{3}{2} \cos^2 \theta - \frac{1}{2} \bigg)\;, \label{equ:ur} \\
u_\theta & = & -3 \frac{x}{r(R - r_{\rm c})} [ 2 \sin (\pi x) + \pi x \cos (\pi x) ] \sin \theta \cos \theta\;. \label{equ:ut}
\end{eqnarray}

\end{appendix}

\end{document}